\documentstyle[11pt,aaspp4,epsf]{article}
\begin{document}

\newcommand{\msun}{\mbox{${\cal M}_\odot$}}
\newcommand{\lsun}{\mbox{${\cal L}_\odot$}}
\newcommand{\kms}{\mbox{km s$^{-1}$}}
\newcommand{\HI}{\mbox{H\,{\sc i}}}
\newcommand{\HeI}{\mbox{He\,{\sc i}}}
\newcommand{\mhi}{\mbox{${\cal M}_{\HIss}$}}
\newcommand{\HIss}{{\rm H\hspace{0.04 em}\scriptscriptstyle I}}
\def\hst{{\it HST}}
\def\rsun{{\rm\,R_\odot}}
\def\etal{{\it et al.}}
\newcommand{\dark}{$\frac{{\cal M}}{L_{V}}$}
\newcommand{\OIII}{\mbox{O\,{\sc iii}}}
\newcommand{\HII}{\mbox{H\,{\sc ii}}}
\newcommand{\NII}{\mbox{N\,{\sc ii}}}
\newcommand{\SII}{\mbox{S\,{\sc ii}}}
\newcommand{\am}[2]{$#1'\,\hspace{-1.7mm}.\hspace{.0mm}#2$}
\newcommand{\as}[2]{$#1''\,\hspace{-1.7mm}.\hspace{.0mm}#2$}

\title{WFPC2 Observations of Compact Star Cluster Nuclei in Low
Luminosity Spiral Galaxies\altaffilmark{1}}
\vskip0.5cm
\author{Lynn D. Matthews\altaffilmark{2,3}}

\author{John S. Gallagher, III\altaffilmark{4}}

\author{John E. Krist\altaffilmark{5}}

\author{Alan M. Watson\altaffilmark{6}}

\author{Christopher J. Burrows\altaffilmark{5}}

\author{Richard E. Griffiths\altaffilmark{7}}

\author{J. Jeff Hester\altaffilmark{8}}

\author{John T. Trauger\altaffilmark{9}}

\author{Gilda E. Ballester\altaffilmark{10}}

\author{John T. Clarke\altaffilmark{10}}

\author{David Crisp\altaffilmark{9}}

\author{Robin W. Evans\altaffilmark{9}}

\author{John G. Hoessel\altaffilmark{4}}

\author{Jon A. Holtzman\altaffilmark{11}}

\author{Jeremy R. Mould\altaffilmark{12}}

\author{Paul A. Scowen\altaffilmark{8}}

\author{Karl R. Stapelfeldt\altaffilmark{9}}

\author{James A. Westphal\altaffilmark{13}}

\altaffiltext{1}{Based on observations with the NASA/ESA {\it Hubble Space
Telescope} obtained at the Space Telescope Science Institute, which is
operated by AURA, Inc. under NASA contract NAS 5-26555}
\altaffiltext{2}{Department of Physics and Astronomy, State University of New 
York at Stony Brook, Stony
Brook, NY 11794-3800}
\altaffiltext{3}{Current address: 
National Radio Astronomy Observatory, 520 Edgemont Road, 
Charlottesville, VA 22903, 
Electronic mail: lmatthew@nrao.edu}
\altaffiltext{4}{University of Wisconsin, Department of Astronomy, 
475 N. Charter St.,
Madison, WI 53706-1582}
\altaffiltext{5}{Space Telescope Science Institute, 3700 San Martin Drive,
Baltimore, MD 21218}
\altaffiltext{6}{Instituto de Astronom{\'\i}a, Universidad 
Nacional Aut\'onoma de 
M\'exico, J. J. Tablada 1006, 58090 Morelia, Michoac\'an, Mexico}
\altaffiltext{7}{Department of Physics, Carnegie Mellon University, 5000 Forbes
Ave, Wean Hall, Pittsburgh, PA 15213}
\altaffiltext{8}{Department of Physics and Astronomy, Arizona State University,
Tyler
Mall, Tempe, AZ 85287}
\altaffiltext{9}{Jet Propulsion Laboratory, MS 183-900, California
Institute of Technology,
    Pasadena, CA 91109}
\altaffiltext{10}{Department of Atmospheric, Oceanic, and Space Sciences,
University of
Michigan, 2455 Hayward, Ann Arbor, MI 48109}
\altaffiltext{11}{Dept. of Astronomy, New Mexico State University, Box 30001
Dept. 4500,
Las Cruces, NM 88003-8001}
\altaffiltext{12}{Mount Stromlo and Siding Spring Observatories, Australian
National
University, Weston Creek Post Office, ACT 2611, Australia}
\altaffiltext{13}{Division of Geological and Planetary Sciences, California
Institute of
Technology, Pasadena, CA 91125}
\singlespace
\tighten
\begin{abstract}
We have used the Wide Field and Planetary Camera~2 (WFPC2) 
aboard the {\it Hubble 
Space Telescope} to obtain $F450W$ and $F814W$ ($B$- and $I$-band)
observations of the compact star cluster
nuclei of the nearby, late-type, 
low-luminosity spiral galaxies NGC~4395, NGC~4242, 
and ESO~359-029. In addition, we analyze archival WFPC2 observations of the 
compact star cluster nucleus of M33.
All of these galaxies are 
structurally diffuse, with
moderately low surface brightnesses and little or no discernible bulge 
component. Here we present a comparative analysis of the structural and 
photometric properties of their nuclei.
NGC~4395 contains a Seyfert~1 nucleus; M33 has some signatures of
weak nuclear activity; the
other two galaxies are not known to be active. All of the nuclei have 
$M_{I}\sim-11$, hence these represent
a little explored low-luminosity extension of the galactic nuclear
activity sequence in a class of host galaxy not 
traditionally associated with galactic nuclear phenomena. 
These kinds of 
compact nuclei appear to be quite common in low luminosity, late-type spirals.

Our Planetary Camera 2 images partially resolve the nuclei of all 
four galaxies. 
A simple model consisting of an isothermal sphere plus a point source
provides a good model for the observed radial intensity distribution
in all cases and permits an exploration of the underlying nuclear
structures and spatial scales. 
Despite their low luminosities, all of the nuclei are
very compact. 
In all cases the  luminosity densities 
are increasing to the resolution
limit of our data at small radii. In spite of having similar size scales and 
luminosities,  the nuclei in 
our sample span a wide range of $B-I$ color. This may be a signature 
of different evolutionary phases in structurally similar nuclei.

The M33 nucleus exhibits complex structure; its isophotes
are elongated, and it has a jet-like component.  
The Seyfert nucleus of NGC~4395  has an 
extremely blue color ($B-I=-$0.16) and  is the most
structurally complex nucleus in our sample. Circularly 
symmetric fits  
to its underlying structure  reveal a distinct bipolar pattern. 
A pair of bright filaments located on one side of the nucleus are  
probably due to [\OIII] emission from gas within a nuclear 
ionization cone.  NGC~4395 appears to
contain an underlying normal star cluster nucleus that is hosting 
activity.  

NGC~4242 shows evidence of a slightly elongated,
bar-shaped feature at its center. The ESO~359-029 nucleus 
appears  relatively symmetric and featureless at the resolution
limit of our data, but it is clearly very compact.

The circumnuclear environments of all four of our program galaxies are
extremely diffuse, have only low to moderate star formation, and
appear to be devoid of large quantities of dust. The central
gravitational potentials of the galaxies are also quite shallow,
making the origin of these types of ``naked'' nuclei problematic.

\end{abstract}


\section{Introduction}
\subsection{Background}
Compact galactic nuclei are a common but still poorly understood
feature of many, if not most spiral galaxies, including our own Milky Way. 
Such nuclei are regions only a few parsecs or less in diameter, but have complex
structures and sometimes produce
energy outputs equal to a significant fraction of the total galaxy
luminosity.  For the present work we define a {\it compact nucleus}
as a non-stellar, point-like light enhancement at or near the 
center of a galaxy whose brightness is in excess of an
extrapolation of the galaxy light profile from the surrounding regions (e.g., 
Phillips \etal\ 1996) and which is dynamically distinct
from the surrounding galaxy disk. Compact nuclei are not simply nuclear 
\HII\ regions, although in some cases, they may 
be one component of a composite nucleus---i.e. they may be
embedded in a nuclear \HII\
region, a nuclear starburst region, or a nuclear disk (e.g., Filippenko
1989; van den Bergh 1995; Ford \etal\ 1997).
Compact galactic nuclei include both ``active galactic nuclei'' 
or ``AGNs'' (i.e.,
QSOs, Seyferts, LINERs), which exhibit characteristic
emission lines and non-stellar spectral energy distributions, as well as 
``compact star cluster nuclei'', whose 
properties can be accounted
for by  dense concentrations of stars.
Compact star cluster nuclei are the densest known stellar systems, 
with even modest examples containing
a mass of more than 10$^6\msun$ within a
few parsecs radius (e.g., Lauer \etal\ 1998; hereafter L98).  Star 
cluster nuclei differ from 
scaled-up versions of 
normal dense star clusters in that they exhibit a wider range 
of stellar ages, quantities of associated dense gas, and sometimes 
evidence for the the existence
of central massive black holes. Possible evolutionary links between AGNs and
compact star cluster nuclei are still widely debated (e.g., 
Norman \& Scoville 1988; Filippenko 1992; 
Williams \& Perry 1994 and references therein).

Compact galactic nuclei can be
difficult to pick out even in nearby galaxies.  Studies from the ground
are limited by  
the effects of seeing and confusion with bright bulges or
regions of enhanced star formation in galaxy centers. Moreover,
in the past, due to the limited 
dynamic range of photographic plates, exposures that revealed the 
outer structures of galaxies often overexposed the nuclear regions, thus
obscuring compact nuclear sources (e.g., van 
den Bergh 1995). 
For some time it was accepted that compact nuclei 
were mainly limited
to dwarf spheroidal galaxies (e.g., Binggeli \etal\ 1984; 
Binggeli \& Cameron 1991) and to elliptical 
galaxies or early-type
spiral bulges in which massive ($10^{6}-10^{7}$\msun) black holes appear to 
be commonplace (e.g., Ford \etal\ 1997; van der Marel 1998 and references 
therein). However, high-quality photographic
plates and high dynamic range CCD images, where the nuclear regions
of galaxies are not ``burned in'', have helped to change our perceptions
on the range in properties of compact nuclei and on the types 
of galaxies that may harbor them (e.g., van den Bergh 1995). 

In particular, it has become evident that true compact nuclei are common in 
low-luminosity, late-type (Scd-Sdm) spiral disks that lack a 
bulge component (i.e., ``extreme 
late-type spirals''). 
Although  the compact nuclei of extreme late-type spirals 
are often relatively faint ($M_{V}\geq\sim -12$), in nearby galaxies they 
become readily detectable even from
the ground under normal ($\sim 1''$) seeing conditions,  
since confusion from bulges or 
brilliant circumnuclear
disks is minimal, and dust obscuration is often low. 

For example, Matthews \& Gallagher (1997) noted  the existence 
of moderate-to-low luminosity compact,
semi-stellar nuclei in 10 of 49 nearby, extreme late-type 
spirals they imaged from the ground. 
In a Snapshot survey with the {\it Hubble
Space Telescope} (\hst), Phillips \etal\ (1996) also found compact
nuclei in a number of nearby, late-type, low-luminosity spirals (see 
also Carollo \etal\ 1997). Compact
nuclei may be common in pure disk galaxies at least to 
moderate redshifts. Sarajedini \etal\ (1996) found that
in a magnitude-limited survey ($I\leq$21.5), 84 of 825 galaxies
contained unresolved nuclear sources. Of these, 57\% of the host galaxies
could be adequately modelled by an exponential disk alone, with 
no bulge component (see also Sarajedini 1996).

The bulk of the compact nuclei in nearby extreme 
late-type spirals appear to be of the ``compact star cluster'' variety (e.g., 
Shields \& Filippenko 1992; Phillips \etal\ 1996). 
Nearby examples include the nucleus of the Local Group Scd spiral
M33 
(e.g., Kormendy \& McClure 1993; hereafter KM; L98) and the nucleus 
of the Sd spiral NGC~7793 (D{\'\i}az \etal\ 1982; 
Shields \& Filippenko 1992). However, in a ground-based spectroscopic 
survey of 43 Scd 
and later nucleated galaxies, Ho (1996)
reported that a surprising
16\% of these objects showed evidence for nuclear activity (i.e., 
Seyfert, LINER, or 
``transition'' nuclei). 

Among the handful of examples of compact nuclei in low-luminosity
disk galaxies that have been studied in detail, there have been 
other surprises.
One of the nearest QSOs (0351+026 
at $z$=0.036) was  found by Bothun \etal\ (1982a,b) to lie within a 
faint ($M_{V}=-18.6$),\footnote{$H_{o}$=75~\kms~Mpc$^{-1}$ is assumed 
throughout this work.} blue, moderately low surface brightness 
disk galaxy.  Moreover, it is interacting with a nucleated, low surface 
brightness, gas-rich galaxy with $M_{V}\sim-17.5$ (Bothun \etal\ 
1982b). 
Kunth \etal\ (1987) were the first to image the galaxy G1200-2038 
associated with a Seyfert~2  nucleus. The host galaxy 
is a very small, faint dwarf galaxy ($D_{25}\approx$6~kpc and 
$M_{V}=-$16.8). The detailed morphology of this
galaxy is uncertain, but it is well represented by a pure 
exponential disk, and its
small physical size and low luminosity are consistent with an extreme
late-type spiral galaxy (cf. Matthews \& Gallagher 1997). In spite of the 
properties of the host galaxy, the nucleus
of G1200-2038 has a luminosity $M_{V}\sim-$16.1, yielding a ratio of 
nuclear to host galaxy luminosity typical of much more luminous AGNs.
Such cases as these may hold important clues as to how 
different ``flavors'' of active nuclei are related, and 
they raise the question of whether
compact nuclei in nearby extreme late-type spirals have special 
characteristics, or perhaps were much more powerful objects in the 
past that have since
run out of fuel.  Furthermore, since  moderate-to-low
luminosity Sd-Sm spirals are the 
most common class of disk galaxy (van der Kruit 1987), if low-level
nuclear activity was commonplace in these objects, it may make 
a significant contribution to the soft X-ray background (e.g., 
Koratkar \etal\ 1995).

However, the majority of nuclei in nearby 
extreme late-type spirals appear to be either
weakly active or non-active, and these galaxies therefore furnish 
examples of the little-explored low-luminosity regime for galactic 
nuclei and nuclear activity.  The investigation of this faint 
end of the galactic
nuclear activity sequence is critical for constraining the origin and 
evolution of compact nuclei and nuclear activity, as well as the physical 
processes that power them, since the existence of compact nuclei in
small diffuse galaxies with weak central potentials and no bulge
component is difficult to explain in light of current models for the
origin of compact galactic nuclei and nuclear activity (see Sect.~6). 
Nonetheless, exploration of these systems is
largely just beginning  (e.g.,
Filippenko \& Sargent 1985; Ho \etal\ 1995a,1997a,c; Koratkar \etal\ 
1995;
Maiolino \& Rieke 1995; Sarajedini \etal\ 1996). 
Still unanswered questions include: Do all star cluster 
nuclei contain
central black holes? Are young stellar populations centrally located
within nuclei? How do structures and spatial 
scales compare 
between active and non-active nuclei? What is the minimum luminosity
for an active nucleus? Because confusion with dust, light from the 
bulge, and bright circumnuclear
material are minimized, detailed studies of the nuclei of nearby extreme 
late-types spirals can help to afford unique 
insights to many of these problems.

By studying the nuclei of extreme late-type spirals, we can also hope 
to further our understanding of the role of the host galaxy in 
provoking and
sustaining nuclear activity, and explore what relationships may exist between
morphology, luminosity, or other global properties of the host galaxy
and its nuclear characteristics. For example
Filippenko \& Sargent (1989) discovered
the faintest Seyfert~1 nucleus yet known ($M_{B}\sim -11$) 
in NGC~4395, one of the objects we explore further in 
the present study. We also investigate the `normal' nuclei 
in the Sd/Sdm galaxies NGC~4242 and ESO~359-029, and analyze
the nearby 
M33 nucleus as a comparison object.

\subsection{Past Observational Limitations and Progress from HST}

Constraining
the physics of compact galactic nuclei requires resolution of their 
structures and measurements of their physical sizes.
However,  since half-light radii
are typically $\leq$10~pc,  even the closest
spiral galaxy nuclei are difficult to resolve
from the ground with conventional observing techniques (e.g., Gallagher 
\etal\ 1982; Nieto \etal\ 1986; Mould \etal\ 1989;
KM). In general, space-based observations with 
image quality of $\leq$\as{0}{1} are needed to procure sufficiently 
detailed
information to advance our understanding of these objects. The first such 
observations were supplied for
M31 by the Stratoscope~II balloon-borne telescope (Light \etal\ 1974).
More extensive data on structures of external 
spiral galaxy nuclei have come 
from ultraviolet (UV) and optical
images obtained with the cameras aboard the {\it HST} that now
achieve $\sim$\as{0}{05} angular resolution. 

Even before its spherical
aberration was corrected, the \hst\ displayed its resolving power by
revealing that M31 has a double nucleus (Lauer \etal\ 1993), showing that
compact nuclei are common in the UV (e.g., Fabbiano \etal\ 1994, 
Maoz \etal\ 1995), and by
providing initial measures of galactic nuclear structures 
(e.g., Crane \etal\ 1993;
Lauer \etal\ 1993,1995; King \etal\ 1995; Ford \etal\ 
1992; Phillips \etal\ 1996). These results have been
confirmed with optical and mid-UV images from the corrected optics of the 
Wide Field Planetary Camera 2 (WFPC2)
and the 
COSTAR-corrected Faint Object Camera (e.g., Colina \etal\ 1997, 
Devereux \etal\ 1997). Aberration-corrected \hst\ 
images of the centers of a few
nearby spirals show that typical galaxy nuclei are often unresolved in the UV,
especially
in galaxies with some evidence for activity.  In the visible, nuclei of spirals
can have complex surroundings (e.g. as in M100),
and range in size from $<$1~pc (e.g., M81, as measured by
Devereux \etal\ 1997)  to core radii of 1.4~pc and 3.7~pc respectively for 
the P1 and P2 double nuclei of M31 (Lauer \etal\  1993; L98). 
 
\subsection{New Goals}
In this paper we present a WFPC2 investigation of the galactic
nuclei of three extreme late-type, Sd-Sdm spiral galaxies. We also 
present an analysis of WFPC2 data of the nucleus of our closest
Scd spiral neighbor, M33, in order to form a comparative framework for our 
analysis.  All of these galaxies
are dominated
by their stellar disks and have little or no bulge component. 
Our objectives are 
to study the optical structures, colors, and spatial scales
of the 
low-luminosity, ``naked'' galactic  nuclei found in these galaxies,
 and to compare these 
results with other compact galactic nuclei.  A preliminary 
version of this work was presented by Matthews \etal\ (1996).

The global properties of our program galaxies 
are described in Table~1. Our first target, NGC~4395, is 
an SAd~III-IV Seyfert~1 galaxy (Filippenko \& Sargent 1989,
Filippenko \etal\ 1993). NGC~4242 (SABd~III) is morphologically
similar to NGC~4395 (cf. Sandage \& Bedke 1994), although slightly
more distant, and  its
optically prominent nucleus does not appear to be active (Ho \etal\ 
1995a).  The
less luminous, nucleated extreme late-type galaxy, ESO~359-029 (Sdm) was 
selected from
the sample of Matthews \& Gallagher (1997; see also Sandage \&
Fomalont 1993) and is one of the lowest 
luminosity disk galaxies ($M_{B}=-15.1$) known to have a nucleus. 
ESO~359-029 does not appear to be active, but a longslit 
spectrum (Matthews 1998) reveals the kinematic 
signature of a compact central mass concentration at the location of
the nucleus.  Lastly, M33 (SAcd~III) 
and its nucleus have similar luminosities to our other targets, 
but because it is nearby, it is well-resolved and has been previously
extensively investigated at multiple wavelengths (see L98). It 
therefore serves 
as an excellent comparison object, a check on the validity of our 
analysis techniques, as well as a guide for the interpretation of 
our results. 

\section{Observations}

Images of NGC~4395, NGC~4242, and ESO~359-029 were obtained 
for this program by the WFPC2 Investigation Definition
Team. The observations are summarized in Table~2.   The nuclei 
of the galaxies 
were observed with the Planetary
Camera~2 (PC2) which
gives a scale of \as{0}{0455} per pixel on an
800$\times$800 pixel Loral CCD.  We used a gain of 7~$e^{-}$ per
data number (DN), and  the CCD was operated at a temperature of 
$-$88$^{\circ}$~C.
More details regarding
WFPC2 can be found in e.g., Trauger \etal\ (1994), Holtzman \etal\ (1995a), 
or Biretta \etal\ (1996). 

Since our program was limited to one
orbit per galaxy, we obtained one short
and two moderate, CR-SPLIT exposures for
each object in the $F450W$ (WFPC2 broad $B$-band) and $F814W$ (WFPC2 broad 
$I$-band)
filters.   The short exposures were 
made to avoid saturated pixels in the bright centers of the nuclei. 
Total exposure times for each object are given in Table~2.
The $F450W$ filter includes the H$\beta$ and [\OIII]
$\lambda\lambda$4959, 5007 
emission
lines, which can be very strong near an AGN. Together, $F450W$ and  
$F814W$ give us a good color baseline.
The data were reduced following the precepts of Holtzman \etal\ (1995a)
and using the approach discussed by Watson \etal\ (1996) for removing cosmic
rays. Calibrations of magnitudes and fluxes are based on 
the system described by Holtzman \etal\ (1995b).

The data for M33 were obtained from the archives of the
Canadian Astronomical Data Centre. These data 
consist of recalibrated WFPC2 PC2 images taken
in the $F555W$ (WFPC2 $V$-band) and $F814W$ filters, and were
pipeline calibrated and combined in IRAF\footnote{IRAF is 
distributed by the
National Optical Astronomy Observatories, which is operated
by the Associated  Universities for Research in Astronomy, Inc.
under cooperative agreement with the National Science Foundation.} to
eliminate cosmic rays. The
archival observations that we used are also summarized in Table~2. 
The $F814W$ images of each of our target nuclei are shown 
in Fig.~1.

\section{A Method for Characterizing the Brightness Profiles of the 
Nuclei}
\subsection{Model Fits}
In order to obtain information about the underlying structure and 
spatial scales of the nuclei of our target galaxies, we have produced 
simple models of 
their light distributions by fitting the data with  combinations of 
point sources plus extended components with simple analytical
forms.  Our approach is motivated 
by the need to extract information from objects which are only 
slightly resolved and to contend with the undersampling of the PC2. 
Both of these effects 
limit the ability to uniquely reconstruct brightness distributions 
of small angular size sources, such as nuclei (e.g., Wildey 
1992; Bouyoucef \etal\  1997). 
For non-active or weakly active nuclei, our models are motivated 
physically by the structure of the
Milky Way nucleus (e.g., Mezger \etal\ 1996) as it would appear at 
larger distances. Our models are also appropriate for star cluster
nuclei hosting AGNs  (e.g., Norman \& Scoville 1988, Perry \& Williams 1993).

Our modelling approach has the advantage of requiring 
a minimum number of free parameters, 
while at the same time yielding measures of the
characteristic sizes of the
nuclei and limits on  their central luminosity
densities. Because our approach does not rely on deconvolution
methods,
we avoid the addition of noise and spurious structures at small radii
that can be introduced by such techniques (e.g., Michard 1996). 
Deconvolution is especially
problematic for undersampled PC2 data, where the solution is strongly
dependent upon the pixel  phase of the adopted PSF (see Hasan \& 
Burrows 1995).
Moreover, if the structures of nuclei are discontinuous at small radii,
and discreet point-like features are present
near their centers, such information
would not necessarily be recovered through deconvolution techniques
(see Sams 1995). Finally, our
modelling approach
allows us to locate off-center or
unsymmetrical features in our nuclei.

The point source contribution for our fits was modelled using 
the Tiny Tim Version 4.1 software
(Krist 1996) to produce a PSF for the PC2
for the appropriate filter, approximate $B-V$ color, 
and the spatial position on the PC2 CCD. The PSF was subsampled by a factor 
of 3 and then interpolated in order  provide finer shifting
and thus improve our ability to align the model PSF with the 
observations. In all cases jitter corrections were 
negligible.

For the extended component of our fits, we considered several different 
analytical forms for the brightness distribution (see Table~3). 
Note that for $\gamma$=2.0, Model~(4) becomes the standard 
Hubble-Reynolds raw (Reynolds 1913; Hubble 1930). For $\gamma$=0.25, 
Model~(6) reduces to the de Vaucouleurs $R^{\frac{1}{4}}$ law (de 
Vaucouleurs 1948), and in the case of $\gamma$=2.0, Model~(8) is
an isothermal sphere (also known as a ``modified Hubble'' or 
``analytical King model''; see Binney \& Tremaine 1987). 
Model~(7) (King 1962; 1966) also reduces to an isothermal sphere in the 
limit $c\equiv log(r_{t}/r_{c})\rightarrow\infty$. As we discuss below, more 
complex models with additional free parameters cannot be adequately 
constrained by the present data.

For each of our target galaxies we attempted fits to the nuclei using 
models of three categories: (a) pure PSFs; (b) pure extended models
of each of the types listed in Table~3; (c) combinations of 
a PSF and extended models. For cases 
(b) and (c), it 
was necessary to convolve the extended component
model with our Tiny Tim model of the intrinsic PSF of the 
optics and to account for the effects of pixel scattering on its 
observed appearance.  We accomplished this transformation by subsampling  
the generated extended model component, convolving 
it with the central 9$\times$9 
pixels of the subsampled Tiny Tim PSF, rebinning to 
normal sampling, and finally, convolving the extended model with the pixel 
scattering kernel. 

\subsection{Fitting Procedure}
Our fitting procedure began with inputting initial guesses for the 
various free parameters in the models. Any or all of the free 
parameters listed in Table~3 were allowed to vary during fitting. In addition, 
we could also allow the peak intensity of the PSF ($I_{PSF}$)
and the 
fractional pixel shift of both the PSF center, $(x_{c},y_{c})_{PSF}$,
and of the extended component center, $(x_{c},y_{c})_{ext}$, to vary.

With our software, fitting was accomplished using a two-dimensional 
non-linear least squares technique based on the CURFIT program of 
Bevington (1969). This method employs the Marquardt (1963) algorithm, 
which combines a gradient search and an analytical solution derived 
from linearizing the fitting function. Individual pixels were weighted 
using Poisson statistics. In all cases, a $\chi^{2}$ 
goodness-of-fit criterion was used to evaluate the 
quality of the models. For each fit, we show a 2-D representation of
goodness-of-fit as: Merit = (Data $-$ Fit)$^{2}~\times$ Weight. These
are a useful tool for comparing the goodness of fit between different nuclei,
since due to normalization uncertainties,
the absolute values of the derived $\chi^{2}$ measures are
difficult to compare directly.  
Outputs from our fits included values for the free parameters for each model 
from Table~3, as well as for $I_{PSF}$, $(x_{c},y_{c})_{ext}$, 
$(x_{c},y_{c})_{PSF}$, the background level, and the integrated fluxes 
of the PSF and the extended component contributions to the final fit. 

To estimate the
sensitivity of our fits to effects of time variability in the PSF and 
camera focus, to mismatches between the true PSF and our model PSF, and 
to the relative strengths of the underlying PSF and extended components,
we generated test images composed of an {\it observed} $F814W$ PSF with 
different focus values added to 
a model
extended component. For the extended model we adopted
a generalized Hubble-Reynolds law [Model~(4)]
with $r_{0}'$=1.0 or 1.5 pixels and $\gamma$=2.0 or 
2.5. We tested focus offsets of up to 5$\mu m$ (i.e.,
1/20$\lambda$), which is the maximum amount of defocus
we expect to see in real images (Biretta \etal\ 1996). We then
attempted to fit these
test images with our fitting software
using a nominal Tiny Tim PSF in perfect focus, plus
a generalized Hubble-Reynolds extended component.  We repeated this 
procedure
on test images where the relative strength of the PSF 
component $I_{PSF}$ to the 
extended component $I_{0}$ ranged from 
10 to 0.35. Regardless of focus, we found the fits 
output by our modelling software to
underestimate the value of $I_{PSF}$ by $\sim$3-5\%. $I_{0}$ was
accurately determined when $I_{PSF}$/$I_{0}$=0.35 but was
overestimated by up to 33\% in the case of  
$I_{PSF}$/$I_{0}$=10. Errors on 
$\gamma$ range from 0-10\%, depending upon both the ratio
$I_{PSF}$/$I_{0}$ 
and the true $\gamma$ of the test image. Finally, for the in-focus 
test images, $r_{0}'$ was systematically underestimated by 2-6\%, while in 
the de-focused test images, $r_{0}'$ was overestimated by 2-26\%, 
depending on $I_{PSF}$/$I_{0}$ and the true underlying 
$r_{0}'$ of the test image. In the 
discussions which follow, we use the above results as partial guidelines for 
estimating the uncertainty of the fit parameters derived from our
modelling technique.

As a second test of the accuracy of the PSF models produced from Tiny Tim,
we attempted to use appropriately scaled versions of the Tiny Tim models
to subtract field stars from our observed image frames. We found the Tiny Tim
models to accurately subtract the stars to within $\sim$2$\sigma$. 
These
residuals are consistent with those expected from the effects of 
Poisson noise and 
the large-angle
scattering in the WFPC2 camera that is not modelled by Tiny Tim. We
adopt the Tiny Tim model PSFs for our nuclear fits rather than using an
observed PSF from each frame, since the observed point sources on our 
frames are of rather low signal-to-noise. In addition, because the PC2 
PSF is position-dependent, the adopted PSF 
must be interpolated to 
match the location of the observed nucleus.
This can be done with much greater accurately 
with a subsampled Tiny Tim PSF than 
with an observed PSF. 

From our tests we deduced the optimal extraction box size for fitting to be
31$\times$31 pixels. This size included the bulk 
of each nucleus' signal, while excluding most of the light from the 
large-angle scattering halo. 
Because M33 is brighter and 
better-resolved (and hence less sensitive to uncertainty in the PSF halo) 
we extended its box size to 61$\times$61 pixels. Modest changes in box 
size caused maximum
changes of a few percent in our derived fit parameters.

\subsubsection{Pure PSF Fits}
To assess how well we had resolved our nuclei, we attempted fits to the
nuclei of all program images using a pure point source (i.e., a Tiny Tim PSF).
We find that {\it none} of 
the nuclei of our target galaxies can be adequately 
fit by a pure PSF. Fig.~2 illustrates this 
for the most distant nucleus in our sample, ESO~359-029.
Clearly a PSF properly scaled to fit the profile wings grossly 
overestimates the central intensity. The mismatch is even more extreme 
in the other 3 cases.
This implies that 
{\it we have partially resolved the nuclei of all four of our program 
galaxies}. For 
NGC~4395 and M33, the 
resolution is  evident even from visual 
inspection of the images. 

\subsubsection{Pure Extended Component Fits}
As a second step,
we attempted to fit all four of our program nuclei
using simple analytical brightness distribution models
of the forms given in Table~3.   We found 
that several of these models [(1), (3), (4), (6), (7), (8), and (9)] 
could adequately fit 
the outer wings of the 
nuclear brightness profiles, but none
was adequate to fit the data over the full range of 
$r$ for any of the nuclei. 
In all instances the models
severely underestimated the flux in the central regions of the nuclei.
Fig.~3 shows this is the case even for the best resolved nucleus in our 
sample, M33. 

\subsubsection{Fits Including Extended Components Plus a PSF}
To better assess the nature of the light distributions of our nuclei 
over the full range of $r$, our final 
series of modelling attempts consisted of 
fitting with combinations of a point source and an 
extended component from Table~3. Our goal was not only 
to try to reproduce the 
brightness profiles of each individual nucleus, but to  
see if we could find a single, unified characterization of 
all of the nuclei that would allow meaningful 
comparisons between them. For this reason, we began with
our best-resolved case, M33.

We first fit the M33 nucleus in the $F814W$ frame 
using combinations of 
a PSF and several of the ``classic'' brightness distributions given in 
Table~3. 
The exponential disk+PSF model fit the wings of the profile adequately, but 
overestimated the peak central brightness. The
Vaucouleurs $R^{\frac{1}{4}}$ law+PSF model was a poor fit for almost the full 
range of $r$, and significantly overestimated the peak brightness 
level. The Hubble-Reynolds law+PSF combination 
roughly fit the shape of the profile for 
most values of $r$, but underestimated the peak central intensity. By 
far the best fit came from the King law+PSF fit, which provided an 
excellent model of the brightness profile for all values of $r$ and 
correctly reproduced the central intensity (Fig.~4). For
this model we found  the 
best fit occurred as $c\rightarrow\infty$---i.e., the best-fit
King model tended toward the case of a simple, isothermal sphere [Model 
(9)].

Using more generalized fitting formulae with additional 
free parameters 
[e.g., Models (4), (6), and (8)]
only marginally improved our fits. One
can trade off between $\gamma$ and the characteristic radii of the models 
to produce a family of fits, none of them 
unique (see also KM). In general, the exponent 
$\gamma$ becomes larger as the characteristic radius ($r_{0}'$, $r_{e}'$, 
or $r_{c}'$) increases.
Thus while they do reproduce the data adequately, 
these sorts of generalized fits have no obvious physical 
meaning, and do not permit useful comparisons between 
the different nuclei. Nonetheless, one important result did emerge:
for the modified King 
model we found $\gamma\rightarrow 2.0$---i.e. this more general form 
again tended  toward the case of a 
isothermal sphere. Together these results suggest that 
an isothermal sphere is an excellent approximation to the outer 
brightness distribution of the M33 nucleus, while at \hst\ resolutions,
the center of the 
nucleus is 
indistinguishable from a point source. Because our 
goal is to fit the observations 
with a useful and reproducible 
model rather than 
exploring all possible model fits (see also KM), we adopt for 
its simplicity and 
minimal number of free parameters, the isothermal sphere+PSF (hereafter 
IS+P)
model as an analysis tool for all of our program nuclei. 

The emergence of the IS+P model for our program nuclei suggests that 
{\it to the resolution limit of our images, the central stellar 
density continues to increase
in all four of our program
nuclei.}  Although we
cannot conclusively rule out other classes of models from our present
data (cf. L98), the IS+P model provides a useful and physically
motivated
characterization of our program nuclei.

Below we present a more detailed analysis of the WFPC2 images of each 
of
our nuclei, including results derived from IS+P
model fitting. For each case, we interpret our findings by 
incorporating 
results from the present work as well as 
prior results from the literature. 

\section{Properties of the Target Nuclei}
\subsection{The M33 Nucleus}
Because of the proximity of M33, its very small bulge, and its
moderate inclination, its nucleus can be effectively studied from the 
ground. Spectra by van den Bergh (1976), Gallagher \etal\ (1982), and 
spectral synthesis models of O'Connell (1983) and Schmidt \etal\ (1990)
show that a range of stellar ages, including stars younger than about 
1~Gyr, exist in the nucleus. Massey \etal\ (1996) deduced from FUV$-$NUV 
colors that the young population has a color  
consistent with a small group of \HeI\ emission stars like that seen 
at the Galactic Center. High resolution ground-based imaging and 
spectroscopy by KM indicated the nucleus is 
very compact with a core radius $r_{c}<$\as{0}{10} ($<$0.4 pc), has a 
stellar velocity dispersion of $\sigma_{\star}$=21$\pm$3~\kms, 
\dark$\sim$0.4, and a maximum central black hole 
mass of 5$\times$10$^{4}$\msun. 

Despite the lack of a very massive 
central
black hole in the M33 nucleus, there are some indications of mild 
activity in the form of [\NII] 
optical emission lines (Rubin \& Ford 1986), possible short- and 
long-term optical variability
(Lyutyi \& Sharov 1990)  and a moderately luminous 
($\sim$10$^{39}$~ergs~s$^{-1}$), 
variable hard X-ray source (Schulman \& Bregman 1995; Dubus 
\etal\ 1997).
Observations of the center of M33 with the WF/PC-1 Planetary 
Camera on \hst\ were presented by Mighell \& Rich (1995). They 
 measured a $V-I$ color-magnitude diagram for surrounding 
field stars which shows a broad red giant branch tip composed 
of stars with ages 
$\geq$1.7~Gyr, younger red supergiants, and a main sequence including 
stars with ages of $\sim$0.1~Gyr. We emphasize that 
M33 cannot be assumed to be simply a
massive version of an ordinary globular cluster because its 
characteristics and environment are much more complex.

The WFPC2 images of the M33 nucleus 
that we present here were also recently analyzed by L98.
These authors  found the M33 nucleus to be 
centrally peaked, which they interpret as a continuous 
increase in mean stellar density towards a 
central `cusp'.  They also noted that 
the nucleus becomes somewhat bluer in color at small radii. 
L98 interpret this combination of properties as possibly being 
due to the presence of binary star merger products in the 
post-core collapse nuclear star cluster, as 
previously suggested by KM. L98 use a deconvolution analysis to derive
a fit to the nuclear light profile of M33. Their
fit is a steep power-law profile for \as{0}{05}$<r<$\as{0}{2} with a
somewhat shallower central cusp or core with
$r_{c}\approx$\as{0}{034}. They place a further limit on the maximum
black hole mass of $M_{BH}<$2$\times10^{4}$\msun, and derive a
central luminosity density of $\sim$5$\times$10$^{6}$\lsun~pc$^{-3}$.

A close  inspection of the WFPC2 images of the 
M33 nucleus reveals several additional
interesting features. 
In agreement with L98, we find the nucleus is clearly elongated, 
especially in the $F555W$ band,
confirming the earlier suggestion of KM. We locate the
major axis at a position angle (PA) of 18$^{\circ}$, in excellent 
agreement with the 
value of PA=17$^{\circ}$ determined by L98. 

At a PA  of 
roughly -20$^{\circ}$ we see evidence of a radial, jet-like feature in both 
the $F555W$ and the $F814W$ frames (Fig.~5a \& b). 
This feature is roughly \as{0}{5} long, and 
contains two bright knots. In the present data, the structure of these
two knots are both consistent with point sources to within errors,
but their location, relative orientation, and colors
are intriguing. The centroids of 
both of these bright spots lie along  
a radial line pointing 
directly toward the center pixel of the nucleus. The
knots have absolute magnitudes (corrected for Galactic extinction) of
$M_{V}$=$-$3.99 (top) and $M_{V}$=$-$5.21 (bottom), 
respectively, and there are no
other discrete sources of similar luminosity in the outer 
parts of the  nucleus or its surroundings. If these are single stars,
they are rather luminous, and it is difficult to explain their highly
disparate colors. The upper source is very red [($V-I$)$_{o}$=1.92]
while the lower source is very blue [($V-I$)$_{o}$=$-$0.22]
(see Fig.~6, discussed below). Even if the sources we see are 
luminous stars,
their alignment with the nucleus center raises the possibility
that they may be associated with a jet or outflow of some sort.
Further investigation of these features
is clearly desirable. If the M33 nucleus 
does harbor a miniature jet, this would provide
evidence that M33 is indeed a low-level AGN. 

We attempted fits to the WFPC2 images of the M33 nucleus using our modelling 
software, as described in Sect.~3.2.3. Our IS+P model
reproduces the light profile of the M33 nucleus quite well, aside 
from small systematic errors due to the slight ellipticity of the nucleus. 
The mismatch between our circularly symmetric models and the slightly 
elliptical nucleus of M33 results in a symmetric residual pattern upon 
subtraction of the model (Fig.~4 \& 7).
Our derived core radius for the IS component of the M33 nucleus 
from the $F814W$ image is $r_{c}$=\as{0}{11}, which is 
in good agreement with the 
measurements of KM and Mighell \& Rich 
(1995), both of whom  found $r_{c}\approx$\as{0}{1}. As 
in the present work, 
KM also included a central point source to reproduce the compact
central source or 
``cusp'' in the brightness profile. Our IS model value for the core
radius is an {\it upper limit} to the physical size of any
core in the M33 nucleus. Our value is larger than the slope break radius  
derived by L98. These authors interpret the unresolved
center of the
nucleus as a density cusp, whose properties were measured via
deconvolution. Our core radius is derived from an
isothermal sphere fit to only  
the {\it resolved} portion
of the underlying nuclear star cluster. 
Our resulting IS+P model 
magnitudes agree well with our aperture photometry (see Sect.~4.7).
For example, the observed $F555W$ magnitude for a 25 pixel circular 
aperture is only 0.01 magnitude fainter than that derived from our model.

We can derive a limit to
the central luminosity density by assuming the flux in 
the model point source in the
unresolved center of the nucleus emanates from a region whose size is
less than one PC2 pixel.  At the
distance of M33, this corresponds to a volume with radius 0.092~pc, and 
the minimum $V$-band luminosity density for the M33 nucleus 
of 3.6$\times$10$^{7}$~\lsun~pc$^{-3}$. The L98 model yields a
lower central density for the M33 nucleus since it assumes 
a continuous model for the nuclear
radial brightness profile.

The color map of the M33 nucleus (Fig.~6) suggests this nucleus has
a blue core, as previously inferred by KM, Mighell \& Rich (1995), and L98.
This result must be interpreted with caution due to the
mismatch between the
{\it HST} PSFs at different wavelengths, but
nonetheless,  the spatial extent of the blue region on our color map
($\sim$3 pixels) is consistent with the slightly bluer nuclear core
measured by L98 in their deconvolved image. From aperture photometry
(see below), we
measure a $V-I$=0.81 within a 3-pixel radius aperture, also
consistent with Fig.~21 of L98.
If real, this central concentration of blue light could represent
a small cluster of young stars (e.g., \HeI\
stars, as suggested by Massey \etal\ 1996), a miniature AGN, or a
single blue supergiant. 
If such discreet sources are present, then the fit of smooth power 
law to the radial intensity distribution becomes uncertain, as discussed 
by Sams (1995). Similar behavior 
is also seen in the Milky Way's nucleus, but in this case it is due to
a  central 
concentration of luminous young stars (Krabbe \etal\ 1995,
Mezger \etal\ 1996). The
central young cluster in the Milky Way nucleus would have a core
radius of about \as{0}{03} at the distance of M33, and therefore would
appear as a blue point source superposed on an older, redder
underlying cluster, similar to what we see in M33. 
We also note that some blue light is present 
in the outskirts
of the M33 nucleus; consistent with the color profile shown in L98, we
see a faint, moderately blue ring surrounding the periphery of the
entire nucleus (see Fig.~6). It is 
clear that even low-mass nuclei 
such as that in M33 are more complicated than simple scaled-up
versions of ordinary star clusters, and the 
radial intensity profile, especially in the central regions, may reflect
more than the stellar  density profile.
Therefore, caution must be exercised in 
assessing the dynamical state of compact nuclei 
from brightness profiles alone (see KM, L98). 

Because M33 is sufficiently resolved, we can also explore its 
brightness distribution using isophotal fits. We determined the run of 
surface brightness with radius for the $F555W$ 10-$s$ exposure using the 
ELLIPSE task in IRAF. To match the slight elongation of the nucleus,
we used a 
fixed ellipticity of $\epsilon$=0.15 at a position angle PA=18$^{\circ}$. The 
background was measured in the regions surrounding 
the nucleus, and a constant background was 
subtracted. The resulting observed radial brightness profile in the 
$F814W$-band is shown in Fig.~8. For comparison, we also fitted the
M33 nuclear brightness profile using circular isophotes. On an
azimuthally-averaged brightness profile, the results appear virtually
indistinguishable from the elliptical fits, indicating the circular
symmetry in the IS+P models should not be a major source of uncertainty.

\subsection{The M33-at-a-Distance Nucleus}

Since our other target nuclei are more distant than M33, our WFPC2 
observations in these cases suffer more severely from limited 
angular resolution. 
However, it is possible to make some comparative measures 
of the size and radial run of intensity in the outer regions of the 
other nuclei. This information provides a basis for  testing the 
hypothesis that other star cluster nuclei in small spirals are
structurally similar to the nucleus of M33. It is useful, before 
undertaking this analysis, to first explore how the observed properties of 
the M33 nucleus would change if it were moved outside of the Local Group.

As a test of the effects of distance on our model fits, we block averaged 
the
$F814W$ image of the M33 nucleus by 5$\times$5 pixels to simulate
its appearance at a distance of about 4~Mpc (hereafter ``M33-at-a-Distance''). 
The WFPC2 $F814W$ image was selected for this 
experiment
because it is less affected by recent star formation or possible 
effects of weak nuclear activity, and so 
provides the best measure of the intrinsic
properties of the underlying nuclear star cluster.  
The main source of uncertainty in this experiment comes from the fact
that in a real galaxy, an unresolved source would 
have its flux distributed over the same number of pixels
regardless of its distance; however in our test image, 
the central point source was binned 5$\times$5.

Our results for the 
IS+P fits to the M33-at-a-Distance nucleus are presented in Table~4
and Fig.~9. These
fits demonstrate a few basic 
points
about the impact of resolution on the fitting process. The fraction of 
the total integrated model flux contained in the PSF increases from 
3.4\% in M33 to 20\% in M33-at-a-Distance. 
The residuals from 
the IS+P fit also become much smaller, demonstrating our lessened 
ability to resolve underlying structure of the nucleus, while at 
the same time the
derived core radius of the extended model component yields a value 
$\sim$1.3 times larger (in parsecs) 
than for the M33 nucleus at its proper distance. 
Thus we cannot measure the form of the central part of distant
nuclei without knowing their radial intensity profile in advance. 

These effects are to be expected. With increasing distance,  
more of the nucleus' light falls within an unresolved point
source, and fewer resolution elements lie across the resolved 
component.
As a result, the extended brightness component of the nucleus
must be separated from the wings of a
strong PSF and cannot be constrained as accurately. Since the PSF 
also effectively removes additional light from the extended component near
the center of the image, the best-fitting extended component will 
necessarily have a shallower brightness gradient at small $r$.
Thus any derived
core radius for the IS component of the 
intensity
model will only be an upper limit to the intrinsic $r_{c}$.

\subsection{The ESO~359-029 Nucleus}

Sandage \& Fomalont (1993) first reported the presence of a point-like 
nucleus in 
ESO~359-029, but classified the host as an Im/dE,N ``mixed morphology'' 
galaxy, and argued that it is gravitationally bound to the Sbc spiral 
NGC~1532. However, CCD imaging by Matthews \& Gallagher (1997) 
emphasized the 
disky appearance of this faint galaxy. They classified ESO~359-029 as an 
Sd, and noted that its disk was
small and diffuse, but very symmetric, 
with a semi-stellar nucleus centered on an ``island'' of slightly
higher surface brightness than the surrounding disk. In addition, a recent 
high-resolution \HI\ 
spectrum by Matthews \etal\ (1998) shows that ESO~359-029 is clearly 
rotationally dominated, that the velocity width reported by Sandage \& 
Fomalont (1993) was underestimated,  and that the \mhi/$L_{V}$ 
ratio of this galaxy (0.41 in solar units) is entirely normal for 
extreme late-type spirals, while being a factor of 10 higher than 
typical dE or dwarf spheroidal systems (e.g., Oosterloo \etal\ 1996 and 
references therein). For these reasons we believe ESO~359-029 is 
best classified as an extreme late-type spiral of type Sd or Sdm.

We note that ESO~359-029 is physically smaller than 
the other 3 galaxies
in the present sample, and shows almost no hint of spiral
structure. Like many extreme late-type spirals, its 
global properties are similar to 
those of an irregular galaxy (Matthews \& Gallagher 1997).
This raises the interesting possibility that at least some nucleated
dE galaxies may form when gas is stripped from an extreme late-type
spiral during a close encounter with another galaxy (cf. Sandage
\& Fomalont 1993). If extreme
late-type spirals were the precursors of dEs, this would eliminate
the difficulty of explaining their origins from irregulars, which are
generally not nucleated (cf. Binggeli 1994).
One way of testing such a scenario is to compare the nuclear
properties of galaxies such as ESO~359-029 with those of dE nuclei
observed at similar spatial resolutions.
One candidate for such a comparison is the Local Group dE NGC~205. Its
nucleus resembles those studied here in terms of its luminosity and size
(Jones \etal\ 1996).

A longslit spectrum  taken by Matthews (1998) revealed 
that ESO~359-029 has detectable H$\alpha$ 
and [\NII] emission
over most of the optical extent of its disk, but no [\SII] was 
detected. The rotation curve of this galaxy is shallow, slowly rising, 
and fairly linear, but shows an abrupt reversal near the nucleus, with 
a semi-amplitude of +25~\kms\ 
and -13~\kms\ on the approaching and receding sides of the nucleus,
respectively. Thus there appears to be a
 kinematic signature of a  compact, massive concentration at the 
center of this tiny spiral.

The results from IS+P model fits to the ESO~359-029 nucleus are 
presented in Table~4 and Fig.~10 \& 11. ESO~359-029 
is satisfactorily fit in both 
the $F450W$ and the $F814W$ bands by the IS+P model. We see only a very 
weak, slightly elongated residual in the $F814W$ frame and essentially no
discernible residuals in the $F450W$ fit.
The core radii we derive for  the resolved component of the
ESO~359-029 nucleus from our IS+P model fits  are 
the largest of the four galaxies in our sample, although ESO~359-029 is 
also the most distant of the four and we emphasize 
our $r_{c}$ values are only upper 
limits (see Sect.~4.2). In 
the blue band 38\% of 
the model flux lies in the point source component of the fit,
suggesting
we have only marginally resolved this nucleus.  
In a manner similar to the case of M33, we derive a minimum central
$B$-band luminosity density for ESO~359-029 from our IS+P model of
2.9$\times$10$^{4}$\lsun~pc$^{-3}$ within a volume of radius 1.10~pc.
Finally, we note that in spite of its faintness (and hence seemingly
comparatively low mass), the nucleus of ESO~359-029 
is clearly very compact. 
The nucleus of ESO~359-029 appears to share basic similarities 
in its properties with the other compact star cluster nuclei in our 
sample, hinting that the properties of
compact star cluster nuclei are at least to some degree
independent of the size and
luminosity of their host galaxies.

\subsection{The NGC~4242 Nucleus}
In the atlas of Sandage \& Bedke 
(1994) the morphological appearance of NGC~4242 is very similar to that of 
NGC~4395. Van den Bergh (1995) drew attention of the prominent 
semi-stellar nucleus visible  in that image. However, unlike the case 
of NGC~4395, there are no spectral signatures of activity in this nucleus. Ho 
\etal\ (1995a) published an optical spectrum of the NGC~4242 nucleus
showing weak emission lines, 
and Heckman (1980) failed to detect a compact nuclear radio source 
down to a limit log$(L_{6 cm})<18.91$~W~Hz$^{-1}$. The lack of 
discernible FIR emission from this nearby spiral by {\it IRAS} and it 
relatively weak global radio continuum flux (Gioia \& Fabbiano 1987) both 
suggest a low rate of global star formation throughout the disk.

Visual inspection of our images of the nucleus of 
NGC~4242 (e.g., Fig.~1c) reveals very faint, extended
``fuzz'' around the main bright nucleus. This is most evident in
the $F814W$ image. Results from IS+P model fitting are presented in 
Table~4
and Fig.~12 \& 13. From this model we derive a minimum $B$-band 
central  luminosity density for
this nucleus of 1.0$\times$10$^{5}$~\lsun~pc$^{-3}$ within a volume of
radius 0.815~pc. 
We find the NGC~4242 nucleus 
is unlike the other nuclei in our 
sample in two important ways. First, this is the only one of the four
nuclei that shows significantly more structure in the $F814W$ images 
than in the $F450W$ image. 
Our IS+P model  provides a good fit to the $F450W$ brightness 
distribution,
but in the $F814W$ band, this model cannot properly
reproduce the flux distribution in the central few pixels,
and leaves an oval-shaped residual 
pattern roughly \as{0}{5}$\times$\as{0}{3} across (see Fig.~13). This
reveals that this nucleus in not circularly symmetric, and that it
appears to contain a miniature bar-like feature near 
its center. We see marginal evidence for a similar feature in the $F814W$ 
image of ESO~359-029, but it is significantly less pronounced. This may 
be due to an intrinsic difference, or to the poorer spatial resolution in the 
ESO~359-029 images.
The second difference between NGC~4242 and the other nuclei 
is that NGC~4242 is the only one of the three
non-active nuclei where the point
source contribution to our model fits is larger in $F814W$ than in the bluer
image.

\subsection{The NGC~4395 Nucleus}

Filippenko \& Sargent (1989) were the first to present evidence that the 
compact nucleus of NGC~4395 is the lowest luminosity Seyfert~1 
known. These authors discovered that the H$\alpha$ emission line 
in the NGC~4395 nucleus has a broad component, and that the [\OIII]
emission is much stronger than that of H$\beta$. Further support 
for the presence of an AGN came from a UV spectrum taken with the Faint 
Object Spectrograph on the {\it HST} by Filippenko \etal\ (1993) which showed
a flat far-UV continuum with strong, high ionization 
emission lines. No P-Cygni profiles were found in the UV, such as 
would normally be present from winds in massive OB stars.
In addition, Filippenko \etal\ found the nucleus to 
be spatially extended in observations made with the original Planetary Camera 
through the $F502N$ filter (a narrow-band \OIII\ filter),
while the continuum 
$F547M$ filter showed only a point source. Sramek (1992) measured a 
non-thermal radio source with about the luminosity of the Cas~A Galactic 
supernova remnant at the position of the NGC~4395 nucleus.
Finally, Lira \& Lawrence (1998) have recently reported the nucleus is
a variable X-ray source. Taken 
together,
the spectral characteristics of the NGC~4395 nucleus appear to be more
consistent with a standard accretion-powered AGN rather than a compact 
starburst.

With our new
WFPC2 data, we have resolved the nucleus 
of NGC~4395.  This nucleus has 
complex internal structure.
In both the $F450W$ and $F814W$ bands, some degree of elongation is 
visible, and a faint halo of irregular ``fuzz'' can be seen 
surrounding the brighter core of the nucleus.

Because the NGC~4395 nucleus is less structurally
complex in the $F814W$ frame, we modelled that image first.
The resolution of the nucleus in $F814W$ is illustrated in Fig.~14, 
where we have subtracted a scaled PSF model to 
approximately  fit the  the central intensity. This leaves behind an 
extended, slightly elliptical residual pattern which contains about 
half of the 
light. A complete fit to the nucleus can be made with good accuracy using our 
IS+P model; this produces a point source and 
extended object centered 
at the same position to within better than 0.1 of a PC2 pixel 
($\approx$5~milliarcsec). As with our other sample nuclei, the 
outer regions of the nucleus of NGC~4395 are reasonably fit by an 
IS+P model (Fig.~16). However, for NGC~4395
an asymmetric, bipolar-like
pattern emerges in the residuals. 
The luminosity of the IS
component of NGC~4395's nucleus in the $I$-band is 
$M_{I}\approx -$10.1 (see Sect.~4.7), similar to the 
resolved nuclear cluster components
in the other galaxies in our sample. The luminosity and 
structural parameters of the NGC~4395 nucleus that we measure from the 
present data are consistent with NGC~4395 containing a normal star 
cluster nucleus which is currently hosting nuclear activity. However, 
final confirmation of this picture will require high angular resolution 
spectra.

The NGC~4395 nucleus looks considerably different 
in the $F450W$ filter.  Two of the central pixels 
of the $F450W$ image were 
saturated in the long exposure, hence these were replaced by values 
derived by scaling from our short exposure (see Sect.~2). 

Visually inspecting the $F450W$ 
image, we see that the  nucleus of NGC~4395
is asymmetric and elongated; it can be 
described as a point source superimposed on an elongated, somewhat 
irregular
structure. We have selected the contour levels in Fig.~16
to attempt to illustrate this point; note the displacement of the point 
source from the outermost isophote. Furthermore, unlike the 
other nuclei, where the centers of the PSF and IS  
components to the fits coincide to within a fraction of a pixel (see Table~4), 
the centroids of the PSF and IS components for the $F450W$ image of 
NGC~4395 are 
offset by 1.2 pixels, and the centroid of the IS
component differs from 
that found in the $F814W$ image by nearly 2  pixels. The luminosity in
the point-source component is considerably higher than would be
expected from the effects of distance alone (based on our
M33-at-a-Distance experiment), arguing that this nucleus truly contains a
very compact source near its center, consistent with other AGNs. 
A fit to the light 
distribution in the $F450W$ frame using our IS+P model 
leaves a bipolar residual pattern similar to that in the $F814W$ 
frame, but much more pronounced, and containing an even larger
fraction of the total nuclear light (Fig.~17). From our IS+P model 
we place a limit on the
minimum $B$-band luminosity density in the NGC~4395 nucleus of
9.6$\times$10$^{6}$~\lsun~pc$^{-3}$ within a volume of radius 0.283~pc.

Further hints of the complex structure of 
the NGC~4395 nucleus can be gleaned from 
an examination of a $F450W-F814W$ color map (Fig.~18). We see overall 
the nucleus is fairly blue, with a compact central source  that appears 
to be somewhat bluer than the surrounding regions. There is also a
short, bright blue arc ($\sim$\as{0}{4} long)
offset just a few pixels from the nucleus center.  This and the prominent 
extended blue plume visible to the right of the nucleus are discussed 
in detail below.

As with the $F814W$ frame, we examine
the residuals left after subtracting 
only the point source component of our model (Fig.~19). From this we see 
evidence of a slight elongation at a PA=255$^{\circ}$ (at ``2 
o'clock'') in the $F450W$ 
image. No 
analogous feature was seen in the $F814W$ image (Fig.~14). The elongated
structure
is only weakly visible, but if the feature
were simply an artifact of an error in the PSF, we would expect to see
a reflected symmetry in this structure, and we do not.
The stronger evidence of the reality of these features is that it
appears to correspond to a feature in our contour plot (Fig.~16) and to 
a distinct blue arc in the color 
map shown in Fig.~18. In the color map, the arc-like
feature is located about \as{0}{25} 
($\sim$3~pc) from the center of the nucleus. The structure 
of this arc and its extremely blue color are consistent with 
it being produced by emission lines from ionized gas (see below). 

These asymmetric blue structures we see in the 
NGC~4395 nucleus are
most readily understood if the light is due to gas rather than
stellar emission;
dynamical time scales within nuclei are extremely short and thus
any azimuthal
irregularities in the distributions of stars will rapidly disappear
in much less than 1~Myr. A stronger confirmation of this picture
comes from the 
consistency of
our observed asymmetry in the NGC~4395 nucleus 
with that found by Filippenko 
\etal\ (1993) in their $F502N$-band 
observations. This agreement suggests that the inner and outer 
filaments are due to regions where the H$\beta$ and [\OIII] emission 
lines are especially intense.  Using the WFPC2 efficiencies tabulated by 
Biretta \etal\ (1996), we find that the blue arc-like feature
produces a flux in [\OIII] of
$\sim4.5\times10^{-14}$~ergs~s$^{-1}$~cm$^{-2}$, 
which is 
about 1/4 to 1/3 of the [\OIII] flux measured from 
ground-based spectra by Ho \etal\ (1997a), and corresponds to a 
luminosity of roughly 1$\times10^{4}$~\lsun. 


The $F450W$ image of NGC~4395 contains a second, large emission plume 
that is clearly resolved from the main nucleus. It 
is approximately aligned with the inner blue arc discussed above. 
This plume also has the blue colors expected for an emission line source, and 
is only faintly visible in the $F814W$ image. The outer plume has a 
wispy, filamentary structure, and
extends about 1$''$ ($\sim$20~pc) in projected distance from the 
nucleus, with its main structure centered near PA=265$^{\circ}$. Near 
PA=270$^{\circ}$, it makes a 90$^{\circ}$ turn and joins onto a fainter 
region at PA=280$^{\circ}$. We detected an
integrated flux from the outer plume is 
44 photons~s$^{-1}$, corresponding to corresponding to a
flux in [\OIII] of 4$\times 10^{-14}$~erg~s$^{-1}$~cm$^{-2}$, or
$L(\OIII)\ge 1.5\times10^{38}$~erg~s$^{-1}\approx 1\times10^{4}$\lsun.
Thus this emission 
line plume, together with the arc discussed above,
contain 1/2-2/3 of the [\OIII] flux found by Ho \etal\ 
(1997a) in their spectra taken with a 1$''\times$4$''$ entrance 
aperture. Since Ho \etal\ (1997a) quote an uncertainty of $\pm$20\% for 
their measurement, our values are in rough agreement.
We therefore believe that these two features comprise the 
bulk of the narrow emission line region around the Seyfert nucleus of 
NGC~4395. 

Our observations of the structure and luminosity of the NGC~4395 nucleus 
and its surroundings are consistent with the behavior of more luminous AGNs.
Normal Seyfert galaxies often show ``ionization cones'' along which 
highly ionized material is found (e.g., Pogge 1989a,b; Wilson \etal\ 
1993; Boksenberg \etal\ 1995;
Mulchaey \etal\ 1996; Schmitt \& Kinney 1996). This 
phenomenon is interpreted as 
blocking or collimation of radiation from the vicinity 
of an accreting black hole. The presence of two 
distinct filaments along approximately the same position angle from the 
nucleus gives strong evidence for the presence of an ionization cone in 
NGC~4395. 
This feature, in combination 
with the spectrum and brightness of the probable [\OIII] line emission,
provide further support to the presence of an AGN within the central parsec 
of the NGC~4395 star cluster nucleus. These factors 
strengthen arguments 
against a stellar origin (e.g., the ``Warmer'' model 
of Terlevich \& Melnick 1985) for
the activity in NGC~4395 (see also Filippenko \etal\ 1993).

\subsection{Comparison of the Structural Properties of the Nuclear
Star Clusters}

The emergence of the isothermal sphere as a suitable approximate model for the 
outer brightness profiles of our nuclei suggests that the nuclei of all of 
our sample galaxies harbor true compact nuclear star clusters, 
not simply small, nuclear \HII\ regions. In contrast to the isothermal
sphere, 
the more general 
King law (King 1962,1966)  
includes a finite value for the cutoff or ``tidal'' radius 
$r_{t}$ (see Eq. 5, Table~3). What does it mean then that our present 
nuclei are better fit as $r_{t}\rightarrow\infty$ (or equivalently, 
$c\rightarrow\infty$) than with the more general King Law? 
One possibility is that we are unable
to measure $r_{t}$  due to our limited field of view and confusion from 
the background light of the galaxy. However, 
in the center of a rigidly rotating galaxy, the tidal radius loses its 
usual physical interpretation, and the nucleus may simply blend into 
the disk, as discussed by KM. The result would then be  
a very slow drop-off in stellar density from a nuclear star 
cluster in galaxies with shallow central gravitational 
potentials; stars will 
only become unbound from the cluster at 
radii where the effects of 
galactic rotation begin to dominate. 

A comparison of the properties we derive for our sample nuclei from 
IS+P model fitting
reveal both
some interesting similarities and some important differences.
The fit results yield two cases 
(M33 and ESO~359-029) where  a larger percentage of the 
integrated model fluxes lie in the point source component 
at bluer wavelengths. Since the intrinsic resolution is slightly 
better at shorter (bluer) wavelengths, this may reflect a real, 
underlying property of these nuclei---i.e., that they contain a compact, blue
central source (or possibly a blue ``cusp'').

In comparing the core radii we derive for the IS component of each of
our nuclei (Table~4), we see some expected effects of distance. 
As predicted 
from our test fits to M33-at-a-Distance, the fraction of
the total flux in the central PSF increases from M33 to NGC~4242 at a
distance of 7.5~Mpc and again for 
ESO~359-029 at 10.2~Mpc. If the overestimation
of the core radius that we measured for M33-at-a-Distance can be 
assumed to scale roughly
linearly with distance, we estimate that the star cluster 
components of all four
of our sample nuclei have similar sizes.
Despite these uncertainties, 
we note that the nuclei for which we 
have both $F450W$ and $F814W$ 
data, our derived core radii from the IS+P models are 
all $\sim$2 times larger in the $F814W$ images than in F450W. This result 
is independent of the differing fraction of the flux in the PSF 
component in the different galaxies. This provides a 
tantalizing hint that our inactive nuclei have underlying structural 
similarities---i.e. an
outlying population of old or intermediate age red stars dominating the 
light at larger radii, 
and an increased blue luminosity contribution 
toward the center. This type of segregation is found in the Milky Way 
nuclear star cluster, although in the Milky Way nucleus the core radius 
of the older, red stars is $\sim$5-10 times that of the inner blue 
population (see Eckart \etal\ 1993, Rieke \& Rieke 1994, Krabbe \etal\ 1995).  
Alternatively, blue nuclear cores could result
from dynamical processes, such as enhanced binary production
leading to excess populations of 
blue straggler stars (KM; L98), or ultra-low luminosity AGNs.

\subsection{Aperture Photometry of the Nuclei}
We established in Sect.~4 that
none of our program nuclei are fully resolved.  Only the nucleus of 
M33, which can be traced over a diameter of about 7$''$ (28~pc), comfortably
exceeds the size of our model of the WFPC2 PSF. However, in all
cases the nuclei stand out from their surroundings on the WFPC2 images 
(Sect.~5),
and thus aperture photometry is relatively straightforward. This is a
major advantage of  {\it HST} data, which typically offer at least ten times
greater angular resolution than optical observations from the ground.

Aperture photometry was carried out using standard software packages in
IRAF. We selected aperture sizes to include approximately 90\% of the
light from each nucleus. For the M33 nucleus we used radii of 50, 75, and
85 pixels (\as{4}{6}, \as{6}{9}, and \as{7}{8} diameters), 30 pixels radius
(\as{2}{76} diameter) for NGC~4242 and NGC~4395, and 20 pixels radius
(\as{1}{84} diameter) for ESO~359-029. Images were background-subtracted prior 
to photometering using background measures obtained from regions
adjacent to the nucleus. Because our galaxies all lack bulges, the 
disk background light was comparatively weak relative to the nucleus.
We derived magnitudes in the 
WFPC2 system as described in Holtzman \etal\ (1995b). These are given 
for each WFPC2 filter $i_{\lambda}$ by:
\begin{equation}
m(i_{\lambda}) = -2.5log(DN~s^{-1}) + 0.748 + ZP(m(i_{\lambda})) 
\end{equation}
\noindent for 
observations made with  GAIN=7. $DN$ is the data number, and the 
filter zero points, $ZP(m(i_{\lambda}))$, are given in Table~9 of 
Holtzman \etal\ (1995b). We adopted observed zero points for the $F555W$ 
and $F814W$ filters and the synthetic zero point for 
$F450W$.\footnote{Technically, magnitudes on the WFPC2 system are defined 
for a point source observed through a \as{0}{5} aperture; there is no 
standard convention for adapting this system to extended sources.}
The resulting WFPC2 instrumental
magnitudes for M33 were converted to $VI$ magnitudes following 
Holtzman \etal\ (1995b), while those of the other galaxies were transformed
to the $BI$ system using new relationships derived by Holtzman (1998):
$$I = m_{F814W} - 0.002(B-I) - 0.009(B-I)^{2}~~~~~(B-I)<1.3$$
$$I = m_{F814W} - 0.004(B-I) - 0.003(B-I)^{2}~~~~~1.3<(B-I)<3$$

$$B = m_{F450W} + 0.108(B-I) - 0.008(B-I)^{2}~~~~~(B-I)<1.3$$
$$B = m_{F450W} + 0.315(B-I) - 0.047(B-I)^{2}~~~~~1.3<(B-I)<3.0$$


\noindent Here $m_{F814W}$ and $m_{F450W}$ refer to the WFPC2 
system
magnitudes given by Eq.~1. Our $B$ and $I$ magnitudes are
presented in Table~5.

Past ground-based measurements can be used to check our results for the M33
nucleus. Photometry by Nieto \& 
Auri\`ere (1982) and by
KM yielded $B=14.5\pm 0.1$ and $B=14.57\pm 
0.07$,
respectively. Combining our observed value of $V=13.80\pm0.05$ with
($B-V$)=0.68 measured by Walker (1964), we find excellent agreement between
our new WFPC2 photometry and ground-based measurements, as well as with 
the WFPC2 measurements of L98. Our $V$-magnitude also agrees within 
errors with the 
value of $V$=13.82 measured by Sharov \& Lyutyi (1988).

Absolute magnitudes in Table~5 are derived from the distances and extinction
values given in Table~1.  
Our sample is not large enough to adequately 
explore possible correlations between host 
galaxy luminosity and the luminosity of the nuclei, although
we do find the most luminous nucleus (that of M33) in the most luminous 
host galaxy and the least luminous nucleus (that of ESO~359-029) in the 
intrinsically faintest host galaxy. However, what is perhaps 
more intriguing is the {\it 
similarity} of the luminosities ($M_{I}\sim$-11$\pm1$) of all 
of the nuclei in spite of the range in luminosities of their host 
galaxies from $M_{V}=-15.1$ to $M_{V}=-18.2$.  
This demonstrates that nuclei of 
similar
luminosities  can be either active or 
non-active and raises the possibility that the nuclei share similar 
origins and long term evolutionary histories.

In spite of the fairly small range in luminosities of our target nuclei, the
spread in their colors is significant. 
The most luminous nucleus, that of M33.  Interestingly,
M33, while having the reddest nucleus in our sample, has the 
bluest nucleus in the Local Group (Sharov \& Lyutyi 1988).
NGC~4242 and ESO~359-029 have intermediate colors, 
both somewhat bluer than typical globular clusters. Finally,
the active nucleus in NGC~4395 stands out because of its
extremely blue optical colors, making it nearly as luminous as the
M33 nucleus in the blue. These blue colors argue against the suggestion 
of Carollo \etal\ (1997)
that faint nuclear star clusters in quiescent disk galaxies are {\it 
old} stellar clusters, but rather hints that 
they could contain a 
mixture of stellar ages. Alternatively, 
binary mergers in old star cluster nuclei could yield a range in 
optical colors (see L98), as could the presence of low luminosity 
AGNs.

 Spectroscopy and spectral synthesis studies 
will be needed to better explore this issue.
Such studies may also yield clues as to why a spread of 
nearly one magnitude in $B-I$ exists among the three non-active nuclei in our 
sample. One possible explanation for the observed color spread 
is that episodic star formation occurs in these types of nuclei (e.g., 
Firmani \& Tutukov 1994). All three galaxies have  similar inclinations
($i\sim 50^{\circ}$), so reddening from extinction 
in the surrounding galactic disks is
unlikely to contribute much to this spread. 

\section{Environments of the Nuclei}

A striking characteristics of the nuclei in our sample is 
the visual appearance of each relative to its respective surrounding 
galactic disk.
In all cases, the nuclei are ``naked''---i.e., they appear 
very distinct from
their circumnuclear environment, and there is little evidence
for significant quantities of 
dust or other material at the centers of these galaxies 
(see Fig.~1).  While to some extent, the nuclei in the present sample
were selected on this basis, we have confirmed this holds even at
space-based resolutions. 
These nuclei thus represent density contrasts of roughly 1000:1 
compared to their surroundings. This
point becomes especially evident in the $F814W$ image of NGC~4395,
where what appear to be two background galaxies can be seen 
through the disk, only tens of parsecs from the nucleus (Fig.~21; for 
contrast with luminous spirals, see the collection of \hst\ nuclear region 
images of Carollo \etal\ 1997).
Furthermore, for all of our sample
galaxies, the continuum light from the galaxy disk 
is only weakly detected during the WFPC2 exposures, and we see no 
hints of spiral structure. While all the 
disks are actively forming stars as evidenced by the presence of 
H$\alpha$ emission, there is little central 
concentration of star formation in any of the galaxies 
(e.g., Wang \etal\ 1997; 
Gallagher \& Matthews unpublished).
These properties of the nuclear regions of our sample galaxies
set  their nuclei apart from those in most other well-studied 
nearby spiral galaxies where the nuclei occur within 
an obvious bulge  component, or the nuclei 
lie at the center of a spiral pattern (e.g., Pi\c{s}mi\c{s} 1987 and 
references therein; Carollo \etal\ 1997; Regan \& Mulchaey 1998) and where
significant quantities of molecular gas and dust
exist in the inner regions of the disk (cf., Carollo \etal\ 1997; 
Devereux \etal\ 1997).

\section{Comments on Possible Formation and Evolutionary Scenarios for 
Compact Star Cluster Nuclei in Extreme Late-Type Spirals}

A possible 
clue to the origin of star cluster nuclei in extreme late-type spirals
comes from the observation of van den Bergh (1995) that nuclei
are {\it not} seen in irregular galaxies, even when these galaxies 
are of similar
luminosities to the objects studied here.  Artyukh \& Ogannisyan (1991) 
also noted that compact nuclear radio sources (indicators of nuclear 
activity) are not generally found in irregular galaxies.
This hints that somehow the 
presence of an organized disk is a requisite factor for the presence of 
a compact nucleus.

The rotations curves   
of  extreme late-type 
spirals tend to be almost linear to the last measured point in the
optical galaxy and have very shallow velocity gradients
(e.g., Goad \& Roberts 1981; Wevers 1984; Matthews 1998). 
This
indicates that the centers of these galaxies are not particularly
``special'' places, and that they have relatively flat 
gravitational potential wells (see also Colin \& Athanassoula 1981). 
This point was also emphasized by KM and
Filippenko \& Sargent (1989) for the cases of M33 and NGC~4395, 
respectively. By contrast, nucleated galaxies 
with luminous bulge components generally 
have steep density profiles with potentials 
pointing sharply towards their centers.
There is some uncertainty whether 
the nuclei of our sample galaxies are even located precisely at the 
dynamical centers of these galaxies (see also Miller \& Smith 1992). 
Minniti \etal\ (1993)  noted that the nucleus of M33 
is displaced by $\sim20''$  from its small ``bulge'' component, and it is also 
displaced from the center of true neighboring disk isophotes (de 
Vaucouleurs \& Freeman 1970; Colin \& Athanassoula 1981).
In addition, the 
velocity centroid of the nucleus of ESO~359-029 appears to be
displaced by $\sim$7~\kms\ 
relative to the galaxy centroid (Matthews 1998). Taken together, the 
global and dynamical 
properties of nucleated extreme late-type spiral galaxies 
raise interesting new questions on the symbiosis between the
nuclei and the parent galaxies and on the origin of these
centralized mass concentrations in otherwise very diffuse disks

Various models proposed for the formation 
of nuclei in early-type galaxies are problematic for the ``naked'' nuclei 
in extreme late-type 
spirals. Capuzzo-Dolcetta (1993) and 
Capuzzo-Dolcetta \& Vignola (1997) have discussed the possibility 
that the nuclei of some early-type galaxies may form from 
globular clusters sinking to the center of the galaxies via dynamical 
friction. However, the efficiency of dynamical friction is greatly 
reduced in galaxies with low central densities (e.g., 
Lin \& Tremaine 1983) making it doubtful that this scenario would be 
efficient in extreme 
late-type spirals. Finally, it appears that at least some 
extreme late-type spirals 
do not contain significant numbers of globular cluster systems (Matthews 
1998). 

Another possibility is that the nuclei formed {\it 
in
situ} from gas infalling onto their centers either due to starbursts 
(e.g., Firmani \& Tutukov 1994) or accretion of primordial gas clumps 
or small gas-rich galaxies (e.g., Loeb \& Rasio 1994). While it seems plausible 
that infalling gas may have provoked episodic 
new star formation and altered the stellar populations of an {\it 
existing} nuclear star cluster in
our galaxies (e.g., Firmani \& Tutukov 1994; Tutukov \& Kr\"ugel 
1995), it is still difficult to explain how gas infalling into shallow 
potentials like those in extreme late-type spirals could have formed the 
compact star clusters at their centers.
Less efficient angular momentum 
transport and weaker gravitational potentials in extreme late-type 
spirals provides a natural explanation for why they appear not to have 
formed supermassive black holes at their centers,  but it cannot 
account for 
the extraordinary compactness of the nuclear star clusters.
This conundrum was pointed out by KM 
with respect to M33.
Our new data reveal that M33 is not an anomaly and that whatever the
conditions necessary for the formation of a compact nucleus in extreme 
late-type spirals, these conditions commonly occur 
in these galaxies.

One possible means of delivering gas to galaxy centers to fuel activity 
or enhanced star formation is via a bar 
(e.g., Shlosman \etal\ 1990;  Ho \etal\ 1995b; Ho 1996). 
Among our sample objects, NGC~4242
is generally classified as weakly barred, while none of the other galaxies,
including M33 have obvious bars. 
Ho (1996) also reported that bars do not appear to enhance 
the probability of finding a compact nucleus in galaxies later than 
Sbc (see also MacKenty 1990; Ho \etal\ 1997b; Mulchaey \& Regan 1997). 
However, it has been pointed out that secondary nuclear bars 
$\sim$1~kpc in length
may be needed to effectively transport material into the central parsec 
region of a galaxy (e.g., Shlosman \etal\ 1989; Friedli 
\& Martinet 1993). Such features have been detected in a few
galaxies (Friedli \etal\ 1996), but we do not see any 
evidence of such bars in our 
program objects. We cannot however rule out that such structures
may have existed in the past and aided the building of the observed 
nuclear star clusters.

Sofue \& Habe (1992) suggested that late-type (Sc and Sd) 
galaxies may have been formed in low density environments and suffered 
at most weak tidal interactions during their evolution. 
This explains the lower typical 
masses of these systems and their lack of large bulges. In this model, 
early-type galaxies formed in denser environments and developed bulges via 
tidally-induced starburst events. If these 
interactions are also responsible for nucleus building and the fueling 
of AGNs (e.g., Loeb \& Rasio 1994), this also could explain 
the lack of evidence for 
supermassive black holes in extreme late-type spirals. 
The low-level activity in NGC~4395 may have been
caused by a very weak interaction or the accretion of a small, gas-rich
satellite that was not enough to cause a significant
alteration in the appearance of the galaxy. In the rare cases, such as 
the galaxy 0351+026, where
extreme late-type spirals are observed to harbor more powerful active 
nuclei (see Sect.~1.1), the tidal effects believed responsible 
for this activity 
(Bothun \etal\ 1982a,b) may irrevocably change the appearance of the 
host galaxy, thus explaining the rarity of such objects in the present 
epoch. An alternative picture is 
that if primordial massive black holes are common in galaxies (e.g., 
Silk \& Rees 1998), the presence of one in a galaxy
may lead to ``puffing'' of the disk (e.g., Faber \etal\ 
1997; Merritt 1998) and the formation of a bulge or spheroid; hence, by 
definition, none of 
these systems will be observed as extreme late-type spiral disks
(cf. Matthews 1998).

M33 is the only case where our present data can help to constrain a
more detailed model of the dynamical evolution of the nucleus.
Because of its small 
velocity dispersion, the predicted 
relaxation time of the M33 nuclear star cluster 
is only 
$t_{relax}\leq$2-3$\times10^{7}$~yr (assuming $r_{c}$=0.1~pc; 
Hernquist \etal\ 1991; KM).
The colors of the M33 nucleus are also 
consistent with most of the stars being older than a relaxation time, so core 
collapse probably has occurred in M33.  
Our new upper bound to the 
the radius of any core in the nucleus of M33 
strengthens this estimate of the relaxation time for simple models of 
core collapse (e.g., Spitzer \& Hart 1971), although  the effects of 
stellar encounters, stellar evolution, and a range of initial 
stellar masses may still create order of magnitude uncertainties in this 
value (KM; L98). Our measured value of the 
approximate core radius for our IS model  
component of the nucleus of M33 
(0.11~pc) also agrees with theoretical predictions of the 
core-collapse scenario (Hernquist \etal\ 1991). 
L98 emphasize that stellar collisions may have particularly 
important consequences in the M33 nucleus, which lacks a supermassive black 
hole to stabilize its structure. 
However, it is likely that the structure of the M33 nucleus is 
more complex than that of an ordinary star cluster, and the detailed 
stellar mass density profile in the center of the nucleus 
is therefore still uncertain. These issues can be most directly 
resolved by spectroscopic observations designed to measure any 
radial trends in stellar content, possible signatures 
of a miniature AGN,  and the nuclear kinematics.

\section{Discussion and Suggestions for Future Work}

We have shown that the nuclei of our program 
galaxies exist in similar, low density 
galactic disk environments and have similar 
luminosities and spatial extents. All of these factors provide evidence 
(but not proof) that the four nuclei have shared common formation 
histories and are comprised of  roughly similar underlying nuclear 
star clusters. We 
postulate that the color and morphological differences we observe
among the various nuclei may be due to relatively short-lived 
evolutionary phases (e.g., episodes of enhanced star formation or 
nuclear activity) superposed on otherwise similar 
nuclear star clusters.

M33's nucleus is the only one of our program objects for which the 
dynamical information exists, hence it is the only case for which a 
definite upper limit may be placed on the mass of a possible black hole 
at its center. We therefore know that any black hole in the nucleus of 
M33 must have a small mass ($M_{BH}<2\times10^{4}$\msun; KM; L98).
Clearly similar measures for the other program nuclei 
are needed to further test the 
hypothesis that the other nuclei in our sample resemble 
the nucleus of  M33. 

A velocity dispersion measure is of particular interest in the case of 
NGC~4395, since it can help to assess whether the activity in NGC~4395 
is due to special properties, or whether it represents a transient 
phase that M33 may have experienced in its past. Filippenko (1992) has 
estimated that if the nucleus of NGC~4395 contains a black hole 
accreting
at roughly the Eddington limit, the black hole mass would 
only be $\sim$100\msun. On the other hand, Filippenko points out that 
the formalism of Wandel \& Yahil (1985) applied to NGC~4395 yields an 
approximate black hole mass of 4$\times$10$^{4}$\msun. This estimate 
assumes that the observed cloud speeds in the NGC~4395 nucleus are 
produced by gravity. This black hole mass estimate is 
nearly 
identical to the upper limit derived for M33 by KM and L98, 
so it is consistent 
with the nuclei of these two galaxies harboring similar central 
massive compact objects.

Further evidence for or against a ``unified'' picture for the nuclei of 
M33 and NGC~4395 could come from X-ray observations. Filippenko \etal\ 
(1993) quoted an unpublished X-ray luminosity of the NGC~4395 nucleus 
of 6.6$\times$10$^{37}$~erg~s$^{-1}$, which is consistent with the 
measurement obtained by Cui (1994). However, Lira \& Lawrence (1998)
report a mean $L_{X}\sim9.7\times10^{37}$~erg~s$^{-1}$ (for D=2.8~Mpc)
with a factor of 2 variability over a 15-day period.
This is roughly a factor of 10
smaller than the X-ray luminosity of the M33 nucleus, but given the 
uncertainty in the distance to NGC~4395, and its slightly smaller total 
nuclear luminosity, it is conceivable that the X-ray luminosity of both 
the M33 and NGC~4395 result from similar mechanisms. However, the
reported X-ray variability occurs on timescales significantly shorter
than the 106-day X-ray variability reported for the M33 nucleus by
Dubus \etal\ (1997).

\section{Summary}
We have analyzed WFPC2 Planetary  Camera broad-band imaging 
observations of the compact star cluster
nuclei of four extreme late-type spiral galaxies: M33, NGC~4395, 
NGC~4242, and ESO~359-029. All of these galaxies have diffuse, moderate-to-low 
surface brightness disks, low luminosities, little or no bulge 
component, and relatively weak central gravitational potentials. The 
nucleus of NGC~4395 is a previously known low-luminosity Seyfert~1; 
M33 has some signs of possible weak activity, while the 
other two nuclei are not known to be active. 
However,  we have confirmed that all four nuclei
appear to be true compact star cluster nuclei rather than simply small
nuclear \HII\ regions.

All of the program nuclei are partially resolved with the Planetary 
Camera 2.
The radial brightness profiles in all four nuclei cases is well fit 
by a combination of an isothermal sphere (IS) and a central
point source (PSF) component, which we refer to as the ``IS+P''
model. Physically this model implies that the structure of the nuclei
may
consist of an underlying star cluster with an abrupt change in
luminosity density near the center. This gradient may be due to 
a power law cusp in the stellar density profile, such as those 
associated with core
collapse (e.g., L98), or could represent the presence of a 
point-like source such as an AGN, a compact group of
young stars, or a single supergiant star. The existence of an unresolved 
center in all of our program objects indicates that the 
luminosity densities in the centers of the
nuclei are increasing to the resolution limit of our data. 

In the case of M33,
the IS+P model provides an
equally good fit to the data as the continuous `nuker' 
power law model of L98. 
Our choice of the simple IS+P model is
physically motivated by the presence of a compact
central light concentration
in the Milky Way's nuclear cluster (seen superposed on a smoother
nuclear star cluster background), and the inferred presence 
of AGNs {\it within} compact nuclear star clusters (e.g., 
Norman \& Scoville 1988). A central subcluster of young stars 
like that in the Milky Way's nucleus  
would be at most marginally
resolved by WFPC2 at the distance of M33, while an AGN 
would be unresolved (see Blandford 1990). 
The IS+P model therefore is on
a firm physical basis for the NGC~4395 nucleus, where the point source 
component represents an extremely compact AGN.
More complex models can provide equally good or slightly 
better fits to the observed nuclear brightness 
maps, but these have more free 
parameters and provide at most 
small improvements in the fit accuracy. 
Furthermore, the more complex models could not be uniquely constrained
and hence do not supply physically insightful information.
While the IS+P model is simplistic, it is also well-matched to the 
degree of resolution we can achieve for galaxy nuclei outside of the 
Local Group.

We 
show that because of the effects of increasing distance, the core radii 
we derive for the underlying nuclear star clusters of our 
program galaxies are only upper limits. We 
estimate the severity of the effects of distance on our measurements 
by producing an artificial ``redshifted'' version of the M33 nucleus. 
From that experiment, we conclude that the sizes of the 
compact star cluster components of all four of 
our program objects are similar to within a factor of 4 and all 
have core radii of less than 1.8~pc in the $F814W$ band.

The luminosities of all four nuclei are similar ($M_{I}\sim 
-11\pm$1) indicating that nuclei with similar luminosities can be either 
active or non-active.
In spite of the modest range 
in optical luminosity, we see a spread of 1.81 
magnitudes in the 
$B-I$ colors of our program nuclei, with the active nucleus of NGC~4395 
being by far the bluest. This spread in color may be 
indicative of different evolutionary phases (e.g., level of nuclear 
activity or recent star formation) in otherwise 
structurally similar underlying nuclei.

In spite of their compact sizes, the nuclei in our sample exhibit 
complex structure.  M33 has slightly elliptical isophotes and shows the 
presence of a jet-like feature. This jet-like feature may be a 
signature
of low-level activity in the M33 nucleus.
NGC~4395 is even more structurally 
complex, especially in the $F450W$ frame, where it appears as 
a point source superimposed on elongated, 
somewhat irregularly-shaped 
isophotes, possibly resulting from ionized gas emission. 
In addition, there 
is a bright arc very near the nucleus ($r \sim$3~pc)  
which is likely due to [\OIII] emission, 
with $L$(\OIII)$\approx1\times10^{4}$\lsun. A larger filamentary 
structure is found along the same line of sight at $r \sim$20~pc, and 
is also likely to be a highly ionized emission region. This pair of  
structures seems to be analogous to the ``ionization cones'' seen in more 
powerful AGNs. These 
features, in combination with the 
luminous blue point source our model fits reveal, 
lend further support to a standard AGN model for the source of 
activity in the NGC~4395 nucleus (see also Filippenko \etal\ 1993).
The similarity of the appearance of the NGC~4395 
nucleus in the $F814W$ filter to the other nuclei in our sample suggests 
that it is a fairly normal star cluster nucleus 
which happens to be presently 
hosting activity. 
Stellar velocity dispersion
measurements will help to confirm or refute this picture.

NGC~4242 is the only one of the four program nuclei that appears more
structurally complex in the $F814W$ frame than in the blue image. Our
IS+P model fits uncover an elongated bar-shaped residual structure at
the center of this nucleus. ESO~359-029 appears relatively symmetric
in our data, but this is the most poorly resolved of our program nuclei.

For the three nuclei for which we have $F450W$ data, we measure an 
IS model component core 
radii that are roughly twice as large in the $F814W$ band as in the $F450W$ 
band.  The blue luminosity therefore 
appears to be centrally concentrated, a symptom that could result 
from a variety of processes. As in the Milky Way's central star cluster, the 
star clusters in our program objects appear to 
have outlying populations of old 
or intermediate-age red stars, and a larger mixture of stellar ages 
in their centers. The presence of miniature AGNs
could also produce this trend, as would 
an excess population of `blue straggler' stars which could form 
via stellar mergers within dense nuclei (KM; L98). 

Various factors make the origin of compact star cluster nuclei in
small spiral galaxies enigmatic.
Our new observations confirm that even when viewed at very high angular 
resolutions,
all four of our program galaxies appear to be 
``naked''---that is they exist 
in extremely diffuse circumnuclear environments with 
a low-density surrounding stellar disk, 
no spiral arms or significant bulge, 
few regions of enhanced star formation, and little indication of dust. 
These
environments are in stark contrast to those typical of brighter 
compact nuclei and more powerful AGNs. In addition, the 
central gravitational 
potentials of the host galaxies of our program nuclei appear 
to be very shallow. Thus it remains unclear how dense, compact
collections of material formed at their centers.
Furthermore, the
morphologies of the host galaxies 
argue against their having been victims of 
major interactions (Matthews \& Gallagher 1997), and their 
low central densities mitigate against dynamical friction being
efficient enough for a star cluster formed elsewhere to migrate to the
galaxy center.

\acknowledgements
LDM has been funded by a graduate internship with the Wide Field and 
Planetary Camera 2 (WFPC2) Investigation Definition Team,  which is
supported at the Jet Propulsion Laboratory (JPL) via the National
Aeronautics and Space Administration (NASA) under contract No.
NAS7-1260.
This work has been carried out as part of the WFPC2 Investigation 
Definition Team 
science programs. We thank D. M. Peterson for comments on an earlier
version of this manuscript.

\newpage

\figcaption{$F814W$-band 
WFPC2 Planetary Camera 2 images of our four program nuclei. 
Field sizes are roughly 
\as{11}{6}$\times$\as{11}{6}. Image orientations are given by the arrows. 
All greyscales are linear.}

\figcaption{Horizontal cut through the ESO~359-029 nucleus in the $F814W$ 
band (solid line) compared with the best-fit Tiny Tim model PSF (dashed 
line). A model PSF scaled to optimize the fit to the wings of the 
nucleus profile greatly overestimates the central intensity, indicating 
the nucleus is not a point source and is partially resolved. Axes are
counts in DN versus pixel number, where x=15 represents the center of
the extraction box (see Text).}

\figcaption{Horizontal cut through the M33 nucleus in the $F814W$ band 
(solid line) compared with the best-fit King Model (dashed line; see Text). A 
King Model reasonably reproduces the wings of the nuclear light 
profile, but severely underestimates the central intensity.}

\figcaption{Horizontal and vertical cuts through the M33 nucleus in the 
$F814W$ band (solid lines) compared with the best IS+P model fit 
(dashed lines).  The four lower panels show 61$\times$61-pixel 
reproductions
of the data, the fit, the merit function (see Text) and the residuals 
left after subtraction of the fit from the data.
Solid black areas indicate 
oversubtraction, white areas indicate undersubtraction of the observed 
flux. The structure in the residuals results from the slight elongation 
of the nucleus compared with the circularly symmetric model. The
$y$-axis has been scaled to a 20-second exposure time.}

\figcaption{Close-ups of the M33 nucleus in the $F555W$ filter
(top) and the $F814W$ filter (bottom). Image sizes 
are roughly \as{5}{8}$\times$\as{5}{8}. The greyscales 
were chosen to emphasize the 
jet-like feature seen in both wavebands at roughly a ``5 o'clock'' 
position. Note that the knots seen in the jet-like structure are 
significantly brighter than any other features surrounding the
nucleus. 
The lower knot is brightest in the $F555W$ 
frame, while the upper knot is brightest in the $F814W$ frame. Note also the 
more pronounced elongation of the nucleus in the $F555W$ image.}

\figcaption{$F555W-F814W$ color map of the M33 nucleus. 
The image is roughly \as{5}{8}$\times$\as{5}{8} across.
The bluest features are seen as black, and the reddest features appear 
white.  Note the compact blue source at the center of the 
nucleus [($V-I$)$_{o}$=0.31], and the presence of both a very blue
source [($V-I$)$_{o}$=$-$0.22] and a very red source [($V-I$)$_{o}$=1.92]
at the position of the
jet-like structure seen in Fig.~5. A faint, moderately blue ring is
also visible surrounding the periphery of the nucleus. The overall
patchy appearance of the color map likely indicates the presence
of dust.}

\figcaption{Same as Fig.~4, but for the $F555W$ band.} 

\figcaption{Azimuthally-averaged representations of the M33 nuclear
radial brightness profile in the $F555W$ band. The solid line indicates the 
observed run of $F555W$ surface brightness (in magnitudes
per square arcsecond) as measured through isophotes with
$\epsilon$=0.15, and plotted as a function of effective radius,
$r_{eff}=\sqrt{\epsilon}a$, where $a$ is the major axis of the elliptical
isophote and $\epsilon$ is the ellipticity. 
The dashed line shows the circularly
symmetric IS+P model intensity distribution, before convolution with
the PSF, where the central point
source is represented as a constant intensity disk with diameter of 1
pixel. The solid squares show 
an azimuthal average of the  circularly symmetric
IS+P model after convolution with the PSF. Note the effect of
PSF convolution in our models is to remove flux from the core and
increase the intensity at larger radii. Lastly, the open triangles
show the deconvolved brightness profile derived by L98 (see their 
Table~5). The excellent agreement
between the observed profile measured through elliptical apertures and
the circularly symmetric IS+P model shows that  our new fit results are not
sensitive to small deviations from circular symmetry. In addition, 
flux is  conserved in our IS+P models. }

\figcaption{Horizontal and vertical cuts through the M33-at-a-Distance 
nucleus in the
$F814W$ band (solid lines) compared with the best  IS+P model fit
(dashed lines). The four lower panels are as in Fig.~4, except that the 
images sizes are 31$\times$31-pixels.  Note the residuals are less 
significant than for M33 (Fig.~4 \& 7).}

\figcaption{Horizontal and vertical cuts through the ESO~359-029 
nucleus in the $F450W$ band (solid lines) compared with the best IS+P 
model fit
(dashed lines).  The four lower panels are as in Fig.~9.}

\figcaption{Same as Fig.~10, but for the $F814W$ filter.}

\figcaption{Horizontal and vertical cuts through the NGC~4242 nucleus 
in the $F450W$ band (solid lines) compared with the best IS+P
model fit
(dashed lines).  The four lower panels are as in Fig.~9.}

\figcaption{Same as Fig.~12, but for the $F814W$ filter. Note the 
prominent bar-shaped structure visible in the residuals.}

\figcaption{31$\times$31 pixel image showing the residuals after 
subtraction of the best-fitting PSF from the NGC~4395 nucleus in the 
$F814W$ image. Note the slight elongation of the residuals, which contain 
roughly half the total nuclear flux.}

\figcaption{Horizontal and vertical cuts through the NGC~4395 nucleus
in the $F814W$ band (solid lines) compared with the best IS+P
model fit
(dashed lines).  The four lower panels are as in Fig.~9. 
Note the bipolar pattern in 
the residuals.}

\figcaption{Contour plot of the NGC~4395 nucleus in the $F450W$ filter. 
Field size is 64$\times$64 pixels.
Contours are spaced at 0.5 magnitude intervals.
The peak surface brightness corresponds to $\mu_{F814W}\approx$13.3
mag~arcsec$^{-2}$. Note the elongation of the nucleus and the 
displacement of the central point source from the outermost isophotes.}

\figcaption{Same as Fig.~16, only for the $F450W$ image. Note the 
stronger intensity of the bipolar residual pattern compared with that 
seen in Fig.~16 for the $F814W$ image.}

\figcaption{31$\times$31 pixel image showing the residuals after
subtraction of the best-fitting PSF from the NGC~4395 nucleus in the
$F450W$ image. Note the slight elongation of the residuals, and the extra
``bump'' at roughly ``2 o'clock'' on the image.}

\figcaption{$F450W-F814W$ color map of the NGC~4395 nucleus. Image size 
is roughly \as{6}{7}$\times$\as{6}{7}. 
Black areas are the bluest, white areas are the reddest. 
The white elongated feature at the lower left-hand corner of the image 
is likely a background galaxy. Note the intense blue color of the plume 
on the left side of the nucleus 
(seen more clearly in Fig.~20), and the small blue arc just to 
the right-hand side 
of the nucleus center.}

\figcaption{$F450W$ image of the NGC~4395 nucleus. Image size is roughly 
\as{11}{6}$\times$\as{11}{6}. The greyscale is set to highlight the 
structure of the emission plume on the righthand side of the nucleus. 
Two of the pixels near the center of the nucleus were saturated in this 
image, and have been replaced by appropriately scaled values from 
our short exposure of this object (see Sect.~2). }

\figcaption{$F814W$ image of the region around the NGC~4395 nucleus. The 
image is roughly \as{23}{2}$\times$\as{23}{2}. The coordinate boxes label 
what appear to be two background galaxies seen through the NGC~4395 
disk. These features are not visible in the $F450W$ image.}

\end{document}